\documentclass[jkps,twoside,twocolumn,fleqn,showpacs,showkeys]{revtex4}
\usepackage{graphicx}
\usepackage{amssymb}
\usepackage{amsmath}
\usepackage{bm}

\def\Jbf{\mbox{\boldmath $J$}}

\def\Mbf{\mbox{\boldmath $M$}}
\def\Hbf{\mbox{\boldmath $H$}}

\begin{document}
\setcounter{page}{0}

\title[]{Enhancement of Curie temperature due to the coupling between 
Fe itinerant electrons and Dy localized electrons in DyFe$_2$Zn$_{20}$} 
\author{Yosikazu \surname{Isikawa}}
\email{isikawa@sci.u-toyama.ac.jp}
\thanks{Tel: 81-76-445-6583}
\author{Toshio \surname{Mizushima}}
\author{Souta \surname{Miyamoto}} 
\author{Keigou \surname{Kumagai}} 
\author{Mako \surname{Nakahara}}, 
\author{Hiroaki \surname{Okuyama}} 
\author{Takashi \surname{Tayama}}
\author{Tomohiko \surname{Kuwai}} 
\affiliation{Graduate School of Science and Engineering, University of Toyama, Gofuku 3190, Toyama, 930-8555, Japan}
\author{Pascal \surname{Lejay}}  
\affiliation{Institut N$\acute e$el, CNRS and Universit$\acute e$ Joseph Fourier, BP166, FR-38042 Grenoble Cedex 9, France}

%\date[]{Received 6 August 2007} 

\begin{abstract}
The temperature dependence of magnetization and specific heat of the ferromagnetic compound DyFe$_2$Zn$_{20}$
has been measured in detail in various magnetic fields.
We have observed anomalous magnetic behavior, i.e., a strong anisotropy at 2 K, 
disappearance of this anisotropy between approximately 30 K and $T_c$, and 
anomalous behavior of the specific heat in the magnetic fields near 20 K.
These anomalous phenomena have been analyzed based on the strong exchange interaction 
between the Fe itinerant electrons and the Dy localized electrons
as well as the crystalline electric field, Zeeman energy, and an usual exchange interaction between two Dy atoms.
The higher $T_c$ of DyFe$_2$Zn$_{20}$ compared with that of DyRu$_2$Zn$_{20}$ is caused by this exchange interaction 
between the Fe and Dy atoms.
\end{abstract}



\maketitle

\section{Introduction}
The series of cubic RX$_2$Zn$_{20}$ has been recently closely examined, where R is a rare earth atom and X is Fe, Co, Ru, etc.\cite{jia1,jia2,jia3,jia4,nasc,tian,wang,isi1}. R is located in the cubic symmetric site surrounded by 16 Zn atoms. The exchange interaction between two R atoms is weak because R atoms are diluted in this compound. In fact, in the case of X = Co or Ru, the Curie temperature is less than 10 K for any R. In the case of the Fe series, however, $T_c$ is significantly enhanced.
It has been suspected that the magnetism of Fe atoms is involved in this high $T_c$.
However, YFe$_2$Zn$_{20}$ is a nonmagnetic material, i.e., the Fe atoms in YFe$_2$Zn$_{20}$ are non-magnetic.
At present, what causes the high $T_c$ of RFe$_2$Zn$_{20}$ is not clear.
We closely examined the magnetization, the magnetic susceptibility and the specific heat of DyFe$_2$Zn$_{20}$ in various magnetic fields, and found experimentally unusual magnetic anisotropies, i.e., the large magneto-crystalline anisotropy at 2 K and a metamagnetic transition in the fields along the [100] direction, 
which disappears in the temperatures above 30 K below $T_c$. 
These anomalous features are analyzed based on the following hamiltonians; the crystalline electric field, the exchange interaction between two R atoms and the exchange interaction between Fe and R atoms. 
We have taken into account the contribution of the 3d electrons of Fe atoms based on the simplified SCR theory.
In this paper, we show that the enhancement of $T_c$, the strong magnetic anisotropy, the metamagnetic behavior at low temperatures, the disappearance of the magnetic anisotropy below $T_c$ down to 30 K, and no anomaly of magnetic specific heat at $T_c$, etc. can be explained by our calculation.

\section{Experimental results}
\begin{figure}[t]
\centering
\includegraphics[width=75mm]{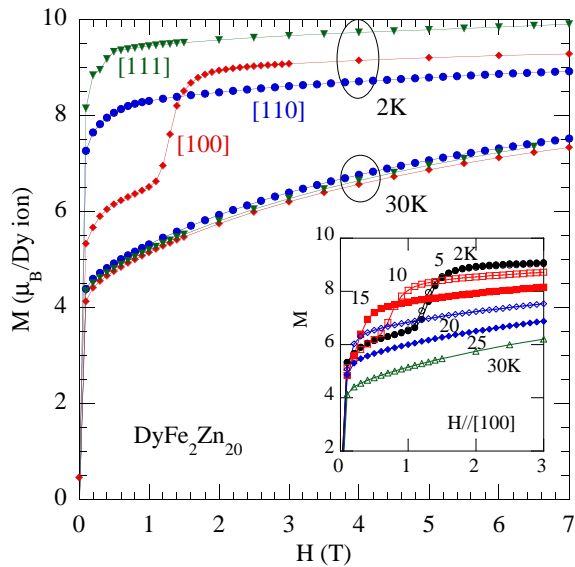}
\caption{Magnetization curves $M(H)$ of DyFe$_2$Zn$_{20}$ at 2 K and 30 K in $H \parallel $  [100],  [110], and [111].
Inset shows the temperature dependence of $M(H)$ in $H \parallel $ [100]. }
\label{MvsH_3}
\end{figure}

Figure \ref{MvsH_3} shows the magnetization curves $M(H)$ in the magnetic fields $H$ along the three principal axes [100], [110] and [111] at 2 and 30 K. $M(H)$ at 2 K is significantly anisotropic, as has been shown by Jia {\it et al.}\cite{jia1}. 
The easy axis of magnetization is the [111] direction. 
At zero field, the values of residual magnetic moments $M$ along the [110] and [100] directions
approximately equal to the projection of $M$ to these directions;
the calculated ratio of $M$ along the  [111], [110] and [100] directions  is 1:$\sqrt{2/3}$:$1/\sqrt{3}$. 
In the figure, a metamagnetic behavior of $M$ is seen in $H$ along the [100] direction in approximately 1.2 T. 
The anisotropic behavior of $M(H)$ disappears at 30 K even though this temperature $T$ is much lower than $T_c$.  
In the inset of Fig. \ref{MvsH_3}, the details of the field dependence of $M(H)$ are shown in $H$ along the [100] direction. 
The field of the metamagnetic transition gradually decreases with increasing temperature, 
and the metamagnetic behavior disappears at 25 K. 
At 15 and 20 K, the metamagnetic behavior is not clearly observed, 
but an unusual $M(H)$ remains in $H$ lower than 0.2 T.  
\begin{figure}[t]
\centering
\includegraphics[width=75mm]{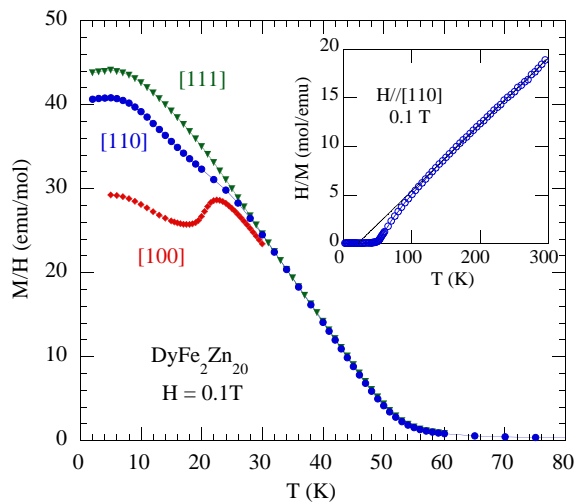}
\caption{Temperature dependence of $M/H$ of DyFe$_2$Zn$_{20}$ in $H \parallel $  [100], [110], and [111]. 
Inset shows the temperature dependence of $H/M$. }
\label{MvsT}
\end{figure}

Figure \ref{MvsT} shows the temperature dependence of $M(T)/H$ in $H$ // [100], [110], and [111]. 
$H$ is applied at 0.1 T. 
From this figure, it is apparent that the strong anisotropy of $M(T)$ at low $T$ rapidly shrinks at approximately 20 K and disappears above 30 K.
The Curie temperature $T_c$ is 53 K, which was determined by $M(T)$ curve in $H = 0.01$ T,
because the temperature change of $M(T)$ at $T_c$ is more clearly observed in low $H$ than that in higher $H$.  
The inset of Fig. 2 shows the temperature dependence of $H/M$, which obeys the Curie-Weiss law. 
The paramagnetic Curie temperature $\theta_p$ is 21.1 K, and the effective moment $\mu_{\rm eff}$ is 10.8 $\mu_{\rm B}$, which is close to the effective moment of the free ion of Dy atom, 10.6 $\mu_{\rm B}$.
The magnetism of DyFe$_2$Zn$_{20}$ seems to be originated from the magnetic ions of rare earth Dy atoms.
However the temperature dependence of $M(T)$ is unusual;
$M$ decreases linearly with increasing $T$, and the temperature change of $M$ at $T_c$ is not clear.
 
The specific heat $C$ of DyFe$_2$Zn$_{20}$ and YFe$_2$Zn$_{20}$ was measured.
The magnetic part $C_{\rm m}$ of DyFe$_2$Zn$_{20}$ was defined as 
$C_{\rm m}=  C_{\rm DyFe_2Zn_{20}} - C_{\rm YFe_2Zn_{20}}$, where
any corrections due to the difference of molar weight
have not been made.
Our $C_{\rm m}(T)$ of DyFe$_2$Zn$_{20}$ was almost the same 
as the one reported by Jia {\it et al.}\cite{jia1}.
The two characteristic features of $C_{\rm m}(T)$, which is not shown in the present paper, 
are (1) a broad peak around 15 K, (2) an unusual and unclear change of $C_{\rm m}$ at $T_c$.

The specific heat $C_{\rm m}$ was measured in $H$ along the three principal axes, 
and we have found anomalous behavior of $C_{\rm m}$ especially below 20 K.
Figure 3 shows the temperature dependence of $C_{\rm m}(T)$ in $H$ along the [100] direction. 
In $H=0$, $C_{\rm m}(T)$ shows only a broad peak with the amplitude of 13 J/mol K at approximately 12 K. 
In 0.2 T, however, a tiny thorn appears at 16 K, 
and then the thorn grows into an apparent peak with increasing $H$, as shown in the figure.
Then the peak abruptly disappears in $H= $ 1.4 T.
The entropy $S_{\rm m}(T)$ reaches to 14.2 J/mol K ($R$ ln 5.6) at 20 K in $H=0$, 
and  decreases to 9.0 J/mol K ($R$ ln 3) at 20 K in $H=8$ T, where $R$ is the gas constant.  
\begin{figure}[t]
\centering
\includegraphics[width=75mm]{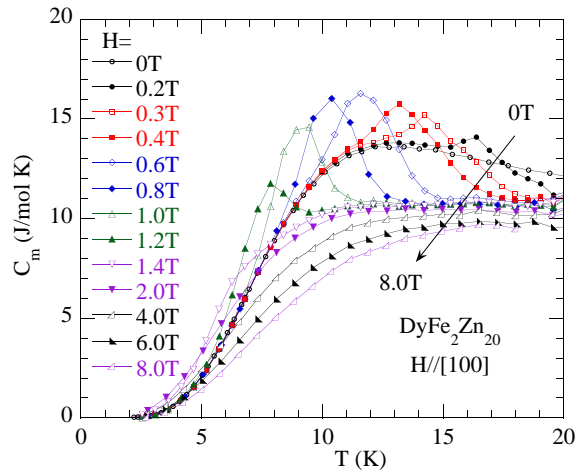}
\caption{Magnetic part of the specific heat $C_{\rm m}(T)$ of DyFe$_2$Zn$_{20}$ in various $H \parallel $ [100].}
\label{fig:3}
\end{figure}
In the case of $H$ along the [111] direction, 
contrary to $C_{\rm m}(T)$ in $H$ // [100],
$C_{\rm m}(T)$ steadily decreases without any anomalies by applying $H$, 
and the broad peak is gradually weakened  with increasing $H$ and disappears in 8 T.

\section{Analyses and discussion}
We analyze the anomalous behavior of $M(T)$, $M(H)$ and $C_{\rm m}(T)$ of DyFe$_2$Zn$_{20}$
taking into account the magnetism of both the Dy and Fe atoms.
First, we estimate the contribution from the Fe atoms based on the experimental results of YFe$_2$Zn$_{20}$,
which is a so-called nearly-ferromagnet.  
Figure 4  and the insets show the inverse magnetic susceptibility $1/\chi(T)$, the magnetization curves $M(H)$ and 
the specific heat divided by $T$, $C(T)/T$.
Note that $M$ is the magnetization per one mole of YFe$_2$Zn$_{20}$, that is, per two Fe atoms.
\begin{figure}[t]
\centering
\includegraphics[width=85mm]{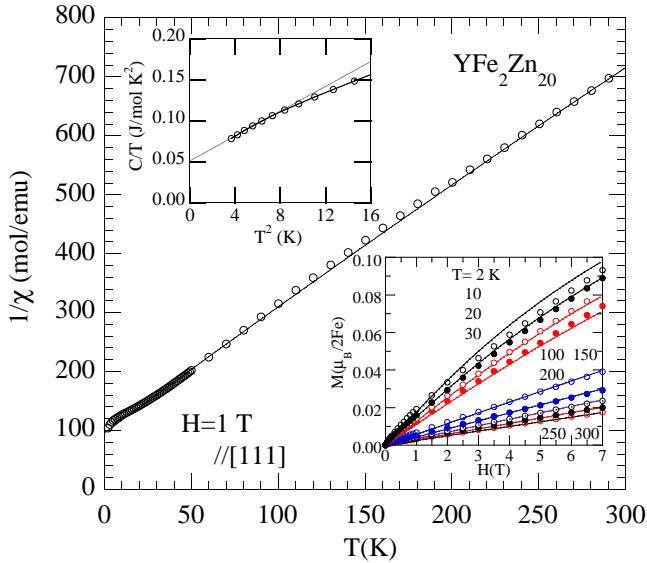}
\caption{Temperature dependence of the inverse magnetic susceptibility of YFe$_2$Zn$_{20}$ in $H \parallel $  [111]. Open circles and the solid line denote the experimental data and theoretical line, respectively. 
Lower inset shows the temperature dependence of $M(H)$ curves at various $T$ in $H \parallel $ [111]. 
Solid lines denote the theoretical lines. 
Upper inset shows the specific heat divided by $T$.
The line in this inset is a guided line for the eyes.}
\label{fig:4}
\end{figure}
$M(H)$ in the inset in Fig. 4 is experimental raw data, 
but the $\chi$ is corrected by subtracting small ferromagnetic impurities less than 0.0005 $\mu_{\rm B}$/2Fe.
The electronic specific-heat coefficient $\gamma_0$ of YFe$_2$Zn$_{20}$ is 0.05 J/mol K, as shown in the inset.
We analyzed the magnetic properties of YFe$_2$Zn$_{20}$ based on the equation,
\begin{eqnarray}
H = \left(\frac{1}{\chi_0}-I+dT^2+b \langle m^2\rangle \right)M_{\rm Fe}+bM_{\rm Fe}^3,
\end{eqnarray}
where $M_{\rm Fe}$ denotes the magnetization of Fe atoms. 
This equation is equivalent to the Stoner model when $\langle m^2\rangle=0$.
The parameter $I$ is a Stoner's enhancement factor, 
and $\langle m^2\rangle$ is a correlation function of magnetization of Fe.
Thus, this equation is equivalent to the SCR theory 
if this term is correctly taken into account\cite{mor1,mor2,taka}.
In this paper, however, to simplify the analysis, we assume that  $\langle m^2\rangle$ is approximately proportional to the temperature $T$,
that is, $b\langle m^2\rangle=cT$.
This assumption can hold in the paramagnetic region, 
because the SCR theory provides the same form as the Curie-Weiss law in this region.
Therefore, experimentally, we can use the equation,
\begin{eqnarray}
H =  (a+cT+dT^2)M_{\rm Fe}+bM_{\rm Fe}^3,
\end{eqnarray}
where $a=1/\chi_0-I$. 
In the limit of $H \rightarrow 0, M_{\rm Fe}  \rightarrow 0$, and thus eq. 2 
corresponds to the Curie-Weiss law (CW) of the  localized model if $d=0$.
The term of $d$ is a correction term from the straight line of the CW law.
These parameters are experimentally estimated as; $a= 96$ mol/emu, $c= 2.18$ mol/emu K, 
$d= 3.7 \times 10 ^{-4}$ mol/emu K$^2$.
The solid line in Fig. 4 is a calculated line using these parameters.
As shown in the inset in Fig. 4, 
at low temperatures in finite magnetic fields, 
$M(H)$ curve is not a straight line. 
The second term in eq. 2 is a correction term to correct the straight line. 
Thus, $b$ is experimentally obtained as 9.19$\times 10^{-5}$ Oe/(emu/mol)$^3$.
The solid lines of $M(H)$ curves in the inset of Fig. 4 are the calculated ones using these parameters.
The parameter $d$ is not effective at low temperatures.
The parameter $\chi_0$ is estimated as 6.85$\times 10^{-4}$ emu/mol
using the experimental value of $\gamma_0$.
The Stoner's enhancement factor $\alpha=\chi_0I$ is 0.93.
\begin{table}[b]
\caption{Characteristic parameters. 
$a$ [mol/emu], $c$ [mol/emu K], $d$ [10$^{-4}$mol/emu K$^2$], 
$b$ [10$^{-5}$Oe/(emu/mol)$^3$],
$\theta_p$ [K], $p_{\rm eff}$ [$\mu_{\rm B}$/Fe], $\alpha$: Stoner's enhancement factor.}
\label{table:1}
\begin{tabular}{ccccccccc}  \hline
sample &  a &    c  &  d    &  b   & $\alpha$ & $\theta_p$ & $p_{\rm eff}$  &\\ \hline
% \#1  &    58 &  2.07 & 0.0   & 8.42 & 0.96   & $-$28.0    & 1.39  & \\ \hline
\#2  &    96 &  2.18 &$-$3.7 & 9.19 & 0.93   & $-$44.1    & 1.36  & \\ \hline
\end{tabular}
\end{table}

Next, we define the following hamiltonian to analyze the anomalous properties of DyFe$_2$Zn$_{20}$,
\begin{eqnarray}
H= H_{\rm CEF}+ H_{\rm exch} + H_{\rm Z} +H_{\rm FeR},
\end{eqnarray}
where $H_{\rm CEF}$ is the crystalline-electric-field (CEF) hamiltonain, $H_{\rm Z}$ the Zeeman hamiltonian, 
and $H_{\rm exch}$ the exchange hamiltonian between the two Dy atoms. 
The former three terms are hamiltonians for Dy atoms in DyFe$_2$Zn$_{20}$.
$H_{\rm CEF}$ is written as $W(\frac{x}{F_4}O_4+\frac{1-|x|}{F_6}O_6)$, 
$H_{\rm Z}$ is $-\Mbf_{\rm R}\Hbf_{\rm ext}$, 
and $H_{\rm exch} = -\Mbf_{\rm R}\Hbf_{\rm mol}$, 
where $\Mbf_{\rm R}$ is the magnetization of Dy atoms, $\Hbf_{\rm ext}$ the external field, 
$\Hbf_{\rm mol}$ the molecular field at Dy atom caused by surrounding Dy atoms,
$\Mbf_{\rm R} = -g_J\mu _{\rm B}\Jbf$, $\Hbf_{\rm mol} = n_{\rm RR}\Mbf_{\rm R}$,
and  $n_{\rm RR}$ the exchange constant between two Dy atoms.
The last term in eq. 3 is an exchange interaction between Dy and Fe atoms, which is written as
\begin{eqnarray}
H_{\rm FeR}= -n_{\rm FeR}  \Mbf_{\rm Fe} \Mbf_{\rm R},
\end{eqnarray}
where $\Mbf_{\rm Fe}$ is the magnetization of Fe atoms, and 
$n_{\rm FeR}$ the exchange constant between Dy and Fe atoms.
Note that the physical quantities of $\Mbf_{\rm R}$, $\Jbf$, $\Hbf_{\rm mol}$, and $\Hbf_{\rm ext}$ are three-dimensional vectors. 
The molecular field at the Dy atom caused by the surrounding Fe atoms is written as
\begin{eqnarray}
n_{\rm FeR}\Mbf_{\rm Fe}=n_{\rm FeR}M_{\rm Fe}
\left(\frac{\Hbf_{\rm ext}+n_{\rm FeR}\Mbf_{\rm R} }{|\Hbf_{\rm ext}+n_{\rm FeR}\Mbf_{\rm R}|}\right),
\end{eqnarray}
where $M_{\rm Fe}=|\Mbf_{\rm Fe}|$, and the term in the parenthesis is a unit vector parallel to $\Mbf_{\rm Fe}$.
In eq. 5, we assumed that 
the magnetic moment of Fe is isotropic, i.e., aligning parallel 
to the direction of $\Hbf_{\rm ext} + n_{\rm FeR}\Mbf_{\rm R}$,
which is the sum of $\Hbf_{\rm ext}$ and the molecular field at the Fe site.
The direction of $\Hbf_{\rm ext}$ is not always parallel to the direction of $\Mbf_{\rm R}$, 
especially in the ferromagnetic temperature region. 
Consequently, 
$M_{\rm Fe}$ obeys the following equation,
\begin{eqnarray}
|\Hbf_{\rm ext} + n_{\rm FeR}\Mbf_{\rm R}|=  (a+cT+dT^2)M_{\rm Fe}+bM_{\rm Fe}^3.
\end{eqnarray}

We summarize the hamiltonian for Dy atom in the frame of the molecular-field approximation as follows: 
\begin{eqnarray}
H &=& H_{\rm CEF}  + g_J\mu_{\rm B}(1+f_{\rm Z}) \Hbf_{\rm ext} \Jbf  \nonumber \\ 
&& -(g_J\mu_{\rm B})^2 (n_{\rm RR}+f_{\rm exch})\langle \Jbf\rangle\Jbf,   
\end{eqnarray}
where
\begin{eqnarray}
f_{\rm Z} &=&  \frac{n_{\rm FeR}M_{\rm Fe}}{|\Hbf_{\rm ext}-g_J\mu_{\rm B}n_{\rm FeR}\langle \Jbf\rangle|},  \\
f_{\rm exch} &=&  \frac{n_{\rm FeR}^2M_{\rm Fe}}{|\Hbf_{\rm ext}-g_J\mu_{\rm B}n_{\rm FeR}\langle \Jbf\rangle|}.  
\end{eqnarray}
The magnetization of Dy atom is calculated by
\begin{eqnarray}
\langle \Jbf\rangle= \frac{{\rm Tr}~ \Jbf\exp(-\beta H)}{{\rm Tr}~ \exp(-\beta H)},
\end{eqnarray}
where $\beta = 1/k_{\rm B}T$.
The specific heat $C$ is calculated by
\begin{eqnarray}
C = N_{\rm A}\frac{\partial}{\partial T}\left (\langle H\rangle -\frac{1}{2}\langle H_{\rm exch}\rangle \right),
\end{eqnarray}
where the second term in the parenthesis is a correction term to correct 
the double-counting of the exchange energy between the two Dy atoms.

There are eight parameters
to fit the experimental $\Mbf(T)$, $\Mbf(\Hbf)$, $C(T)$, and $C(\Hbf)$ in this calculation:
 $x,~W,~n_{\rm RR},~n_{\rm FeR}, a,~ c,~d$ and $b$.
However, the parameters $a,~c,~d$ and $b$ are fixed as in YFe$_2$Zn$_{20}$.
Moreover, the parameters $x,~W$ and $n_{\rm RR}$ are estimated by adjusting  
the magnetic crystalline anisotropy and $T_c$ for  DyRu$_2$Zn$_{20}$, 
which is the compound without magnetic Fe atoms, where $T_c$ of DyRu$_2$Zn$_{20}$ is 4 K.
The final parameter is $n_{\rm FeR}$, which is determined by
adjusting $T_c$ of DyFe$_2$Zn$_{20}$.
The eight fitting parameters seem to be easily found. 

\par
However, difficulties of the calculation exist 
in the iteration procedures 
when we solve simultaneously  both $\Mbf_{\rm R}$ of Dy atoms and $\Mbf_{\rm Fe}$ of Fe atoms
using eqs. 6 and 10. 
In this paper, we have adopted a two-step procedure to solve these equations.
First, we ignore the anisotropy of $\Mbf_{\rm R}$.
When $H_{\rm CEF}$ is excluded from the total hamiltonian,
the directions of $\Mbf_{\rm Fe}$ and $\Mbf_{\rm R}$ are parallel to $\Hbf_{\rm ext}$ in the case of positive $n_{\rm FeR}$.
$M_{\rm R}$ is easily expressed using the Brillouin function $B_J(x)$ as
\begin{eqnarray}
\langle J_z\rangle = -JB_J(x),
\end{eqnarray}
where
\begin{eqnarray}
x =
\left(\frac{g_J\mu_{\rm B}J(H+n_{\rm RR}\langle J_z\rangle +n_{\rm FeR}M_{\rm Fe})}{k_{\rm B}T}\right).
\end{eqnarray}
In the case of negative $n_{\rm FeR}$, 
the directions of $\Mbf_{\rm Fe}$ and $\Mbf_{\rm R}$ are not always parallel to $\Hbf_{\rm ext}$.
However, we used eq. 12 for both cases for simplicity.
Using eqs. 6 and 12, $M_{\rm Fe}$ and $\langle J_z\rangle$ are easily calculated. 
As the results of this step,
the contribution of $M_{\rm Fe}$ is found to be  approximately 4 \% of $M_{\rm R}$.
Then, as the second step of the calculation, 
the anisotropic $\Jbf$ of Dy atom is re-calculated based on eq. 10 
by the iteration procedure.
In this step, we fixed $M_{\rm Fe}$ at the value calculated in the first step.

Figures 5, 6 and 7 show the calculated results of $M_{\rm R}(H)$, $M_{\rm R}(T)$ and $C(T)$, respectively. 
The calculated $M$ in these figures is $M_{\rm R}$, because $M_{\rm Fe}$ is presumably smaller than $M_{\rm R}$. 
The CEF parameters and the exchange interaction constants are listed in Table 2.
\begin{figure}[t]
\centering
\includegraphics[width=75mm]{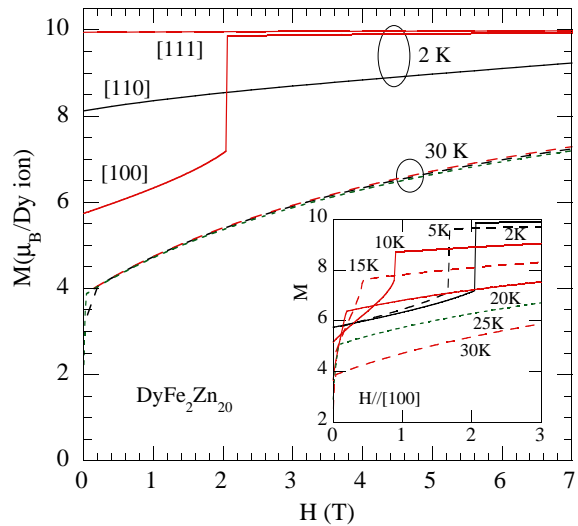}
\caption{Calculated $M(H)$ curves at 2 and 30 K in $H \parallel $ [100], [110], and [111].
Inset shows the calculated $M(H)$ curves at various $T$ in $H \parallel $  [100].}
\label{fig:5}
\end{figure}
\begin{figure}[t]
\centering
\includegraphics[width=75mm]{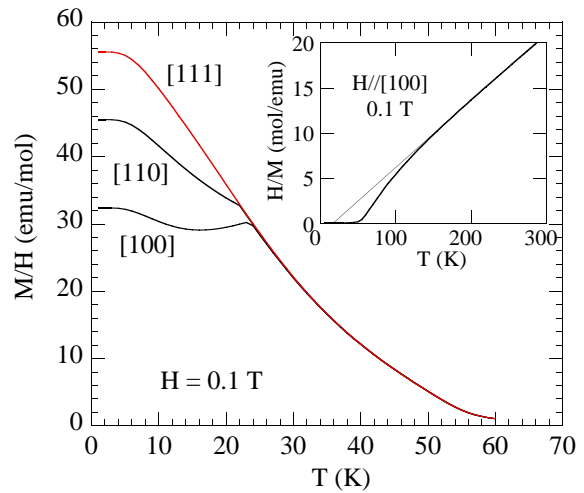}
\caption{Calculated temperature dependence of $M/H$ in $H \parallel $ [100], [110] and [111].
Inset shows the calculated $H/M$ in $H \parallel $ [100]. The thin line in the inset is a guided line for the eyes.}
\label{fig6}
\end{figure}
\begin{figure}[t]
\centering
\includegraphics[width=75mm]{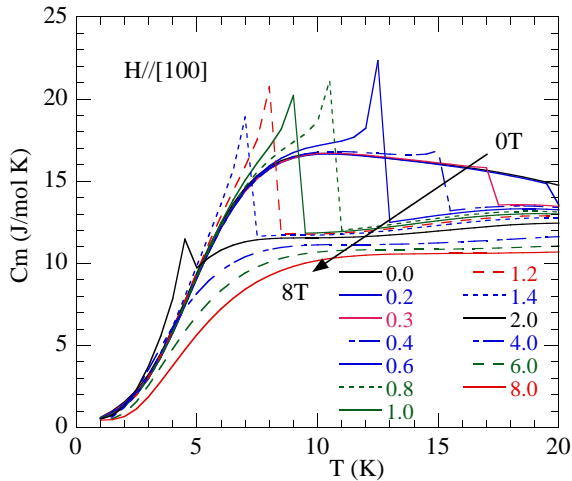}
\caption{Calculated  specific heat $C_{\rm m}(T)$ in various $H \parallel $ [100].}
\label{fig:7}
\end{figure}
\begin{table}[b]
\caption{CEF parameters, $W$ [K] and $x$, and the exchange constants between R-R and R-Fe, 
$n_{\rm RR}$ [T/$\mu_{\rm B}$] and $n_{\rm FeR}$ [T/$\mu_{\rm B}$].}
\label{table:2}
\begin{tabular}{ccccc}  \hline
 $W$     & $x$    & $n_{\rm RR}$ & $n_{\rm FeR}$ \\ \hline
$-0.036$ & $-0.2$ & 0.188        & $-14.8$       \\ \hline
\end{tabular}
\end{table}
The energy splitting by CEF is considerably small; the ground state is  
$\Gamma_8=-7.5$ K, the first excited state is 
$\Gamma_8=-3.5$ K, and the overall energy splitting is 16.3 K.

The calculated results of $M(H)$ and $M(T)$ are in good agreement with the experimental results in Figs. 1 and 2 
both qualitatively and quantitatively. 
The calculated results represent the following characteristic features of DyFe$_2$Zn$_{20}$: 
the calculated anisotropy $M(H)$ at 2 K coincides with the experimental one, 
the easy axis of magnetization is parallel to [111], 
the metamagnetic transition occurs in $H$ along the [100] direction in approximately 2 T at 2 K, 
the metamagnetic transition field decreases with increasing $T$ and goes down to zero near 20 K, 
the magnetic anisotropy disappears for all the principal axes above 30 K, 
Concerning $C(T)$ in $H$ along the [100] direction, the calculated results reproduce well the experimental features:
$C(T)$ in $H=0$ makes a broad peak at approximately 10 K without any anomalous thorn,
a small anomalous thorn appears by applying $H$ along the [100] direction 
and it grows into an apparent peak with increasing $H$ and then disappears above 3 T.
In the cases of $H$ along the [110] and [111] directions, 
we could also analyze the experimental results of $M$ and $C$ by our calculation.  

We summarize the experimental and calculated results of DyFe$_2$Zn$_{20}$.
DyFe$_2$Zn$_{20}$ is a ferrimagnet with two magnetic moments of Dy and Fe atoms. 
The former ion has 10 $\mu_{\rm B}$/Dy, the latter 0.4 $\mu_{\rm _B}$/2Fe.
This small magnetic moment of Fe atom  has been currently inferred 
from the M$\ddot{\rm o}$ssbauer experiments\cite{tamu}. 
The overall splitting of CEF is 16 K. 
The exchange interaction $n_{\rm RR}$ between the two Dy ions is extremely weak compared with the exchange interaction $n_{\rm FeR}$ between the Dy and Fe atoms.
Thus the $T_c$ of DyFe$_2$Zn$_{20}$ is enhanced by this $n_{\rm FeR}$, because 
it is 80 times larger in magnitude than $n_{\rm RR}$ in DyFe$_2$Zn$_{20}$.
This small $n_{\rm RR}$ is caused by the distance between Dy ions in DyFe$_2$Zn$_{20}$, 
and the large value of $n_{\rm FeR}$ is presumably 
due to the large overlapping of the wave function between the itinerant 3d and the localized 4f electrons.

\vskip 5mm
This study is partly supported by a Grant in Aid for Scientific Research (No. 21540356) from the Japan Society for the Promotion of Science.

\vfill
\end{document}